\documentclass[10pt,a4paper]{article}

\usepackage{lipsum}
\usepackage{url}
\usepackage{amsfonts,amsmath,amssymb,amsthm}
\usepackage{stmaryrd}
\usepackage[nochapters]{classicthesis}
\usepackage[shortlabels]{enumitem}
\usepackage{xurl}
\usepackage{hyperref}
\usepackage{graphicx}
\usepackage{longtable}
\usepackage{changepage}
\usepackage{tabularx}
\usepackage{booktabs}

\newcommand\blfootnote[1]{%
  \begingroup
  \renewcommand\thefootnote{}\footnote{#1}%
  \addtocounter{footnote}{-1}%
  \endgroup
}

\begin{document}

\title{\spacedallcaps{ASTRA: AI Safety, Trust, \& Risk Assessment}}
\author{
Pranav Aggarwal$^{1}$, 
Ananya Basotia$^{1}$, 
Debayan Gupta$^{1}$, \\ 
Rahul Kulkarni$^{3}$, 
Shalini Kapoor$^{2}$, 
Kashyap J.$^{1}$, \\
A. Mukundan$^{1}$, 
Aishwarya Pokhriyal$^{1}$, \\
Anirban Sen$^{1}$, 
Aryan Shah$^{1}$, 
Aalok Thakkar$^{1}$
\\\\
$^{1}$\spacedlowsmallcaps{Ashoka University} \quad 
$^{2}$\spacedlowsmallcaps{EkStep Foundation} \quad
$^{3}$\spacedlowsmallcaps{DoNew} \\
}
\date{\spacedlowsmallcaps{February 2026}}
    
    \maketitle

    \section{Introduction}

India\blfootnote{This work is generously supported by EkStep Foundation’s AI and Data Adoption Fellowship.} is a nation of profound complexity that frequently defies conventional socio-technical classification. With a population nearing 1.5 billion, hundreds of distinct languages, and a biometric identity infrastructure achieving near-universal coverage, India has demonstrated a unique capacity for large-scale digital transformation. Exemplars such as Aadhaar and the Unified Payments Interface (UPI) underscore a national aptitude for deploying Digital Public Goods (DPGs) at scale. This rapid digitalization has been a cornerstone of India’s developmental trajectory, contributing significantly to the reduction of poverty from 30\% to just over 10\% within a single decade.

The current inflection point in Artificial Intelligence (AI) and Machine Learning (ML) offers transformative potential for the Indian citizenry. When coupled with the prospects of significant private investment and national wealth creation, the adoption of AI becomes a strategic imperative that policymakers cannot ignore. However, the governance of a field as nascent and protean as AI, without stifling innovation at this critical developmental juncture, presents a formidable challenge. We must ensure that AI serves as a catalyst for domestic empowerment rather than a new conduit for techno-colonial exploitation that primarily extracts value for the Global North.

Despite their sophisticated capabilities, modern ML models, particularly Large Language Models (LLMs), possess inherent vulnerabilities regarding reliability and fairness. Unlike classical physical engineering, ML systems generally lack formal error bounds or deterministic correctness guarantees. Their performance is typically validated on static distributions similar to training data, which often fails to generalize in dynamic, real-world environments. Furthermore, LLMs are mathematically optimized for linguistic coherence rather than factual veracity, leading to \emph{hallucinations}, i.e. outputs that are syntactically persuasive yet factually erroneous. Crucially, the failure modes of ML models diverge from human cognition; machines and humans encode complex concepts through fundamentally different architectures. Consequently, inputs that appear semantically proximate to humans may be processed as disparate by a model. Theoretical constraints also demonstrate that achieving parity of outcomes across diverse demographic groups is mathematically impossible unless the underlying data distributions satisfy specific, often unattainable, criteria.

Deploying large-scale ML models in public-facing roles thus entails significant risks that are often obscured by the \emph{AI hype.} These harms are socially expensive; they can entrench historical biases, obfuscate institutional accountability, and subtly reshape public perception. Given the specialized jargon and technical opacity surrounding these risks, decision-makers across government and industry often possess a fragmented understanding of the stakes. This leads to a superficial engagement with the technology, where terms like ‘hallucination,’ ‘blackbox,’ and ‘explainability’ are used as buzzwords rather than rigorous analytical categories.

\subsection{Defining Risk}

In a foundational sense, ``risk'' represents the probability of an undesirable outcome, characterized by the product of its likelihood and its severity. In the context of AI, negative outcomes typically involve the infringement of fundamental rights or the violation of statutory obligations. While this concept is well-established in actuarial science (ranging from physical safety to constitutional morality) AI risks present a unique epistemological challenge: while we can classify potential harms, we often cannot accurately estimate their probability. Unlike actuarial risks like vehicular accidents, where historical frequencies provide a statistically valid basis for prediction, AI failures are often black swan events or emergent behaviours. However, while probability remains elusive, the potential severity of an outcome can, and must, be rigorously assessed.

\subsubsection{AI Safety Risks (ASR)}

In alignment with evolving global standards, we define AI Safety Risks (ASRs) as hazards emerging from design flaws that cause a system to deviate from its intended functional parameters. It is vital to recognize that while ASRs are rooted in design, they manifest across the entire lifecycle: development, deployment, and usage. By the time a harm is realized in a production environment, the causal conditions were likely established during the architectural phase.

For example, risks of bias or socio-economic exclusion resulting from skewed training sets or flawed modeling assumptions are quintessential ASRs. Conversely, a system failure caused by physical infrastructure destruction is an operational risk, not an ASR, as the failure is exogenous to the system’s design. However, when a user leverages an AI to generate malicious deepfakes, the harm manifests at the point of usage, but the root cause is a design deficiency, specifically, the absence of robust safety guardrails. Thus, it remains an ASR. Notably, intended harms such as the calculated use of autonomous weapons in warfare fall outside the scope of ASR, as the system is performing exactly as designed.

Distinguishing ASRs requires a nuanced understanding of causality. To assist practitioners, we propose a practical heuristic: \textit{a risk is classified as an ASR if it can conceivably be mitigated by the deployment of a new version of the AI.} If a more constrained architecture or a more representative training regime can eliminate the harm, it is a safety risk. This \emph{remediability heuristic} allows laypersons to distinguish between inherent model risks and external operational failures.

Furthermore, the granularity of ASRs is deeply contingent upon the legal and socioeconomic context. A primary example is caste-based bias, a risk specific to the Indian subcontinent that is absent from Western taxonomies. Such nuances manifest in privacy concerns as well; while Western frameworks focus on individual identifiers, many Indians use initials to mask caste identity, creating a unique challenge for bias detection and data privacy in a pluralistic society.

\subsubsection{Systemic Risks around AI}

Broader societal risks including cognitive atrophy, labor market disruption, or digital addiction fall into a separate category. These risks are ill-defined, systemic, and cannot be solved through technical iteration alone. While these represent some of the most existential threats posed by AI, they are beyond the technical scope of this document, which focuses on remediable safety risks.

\subsection{The AI Safety Framework}

An AI Safety Framework serves as the architectural blueprint for institutionalizing trust. It encompasses the principles, governance structures, and feedback loops that manage the AI lifecycle, from risk identification and mitigation to continuous monitoring. A robust framework must produce interoperable standards across industries. In the current landscape, where innovation outpaces legislation, assurance becomes the primary mechanism for developers to signal that a system is aligned with societal values and legal mandates. These standards provide a common metric for evaluating impact and managing systemic liabilities.

\subsection{The Indian Context}

While several international AI risk taxonomies exist, they frequently suffer from \emph{contextual blindness} when applied to India’s socio-technical reality. Relying on external standards risks a form of \emph{algorithmic colonization,} where Indian systems are audited against benchmarks that do not account for local linguistic or infrastructural constraints.

In India, the state views AI primarily as an instrument of national development. Unlike the market-centric narratives of the Global North, India’s objective is to \emph{transform lives} through Digital Public Goods. This developmental lens necessitates three critical shifts in risk taxonomy:

\begin{itemize}
\item \textbf{Empirical Grounding:} Policy must be rooted in the current functional reality of AI and the domestic socio-economic landscape. We must resist \emph{venture-capital optimism} and focus on verifiable capabilities.
\item \textbf{Opportunity Cost as Risk:} Risk should be measured not only by the presence of harm but by the failure to achieve public-interest objectives. If an AI tool for agriculture benefits large-scale farmers while marginalizing smallholders due to connectivity issues, the \emph{risk} is the failure of the developmental mandate.
\item \textbf{Sovereignty and Autonomy:} Taxonomies must track supply-chain dependencies. Heavy reliance on foreign APIs for critical identity or payment infrastructure introduces a fragility that Western narratives focused on efficiency and profit often overlook.
\end{itemize}

Standard AI models are typically trained on datasets from the Global North, reflecting cultural and institutional realities that differ from India’s pluralistic society. Many off-the-shelf frameworks assume high levels of digital literacy and uniform data access. In the Indian context, a taxonomy must account for the exclusion of vernacular speakers from welfare systems or the bias inherent in models that do not recognize under-represented regional dialects.

Furthermore, India’s economy is characterized by a massive informal sector. AI applications here often function in high-stakes public sectors such as land records, healthcare in low-resource settings, and welfare distribution rather than purely commercial fintech. Consequently, risk typologies must prioritize digital exclusion and regional inequity. While Western frameworks might focus on autonomous vehicles, the salient risks in India involve the mis-targeting of life-critical entitlements.

Institutional maturity is also a factor. While the IndiaAI Mission is a significant step forward, the empirical database for documented AI harms in India is currently insufficient. As noted recently, India lacks a risk classification framework based on empirical evidence. Without this, frameworks remain purely theoretical. Moreover, the cost of compliance must be calibrated to India’s unique ecosystem; a framework requiring expensive, hyper-specialized audit infrastructure may inadvertently stifle the very startups and public agencies driving social good.

In conclusion, while global frameworks provide a starting point, they require rigorous contextualization to be effective in India. This whitepaper represents the first step in this journey: risk identification. We propose an ASR database tailored to DPGs deployed across Indian sectors. This version focuses specifically on the domains of Education and Financial Lending, establishing a scalable foundation for a comprehensive Indian AI Safety Framework.
    \section{Related Work}

‘Concrete Problems in AI Safety’ (2016) addresses the problem of accidents in machine learning systems, defining an accident as unintended and harmful behavior emerging from poor design of real-world AI systems. It outlines five concrete risks: negative side-effects, reward hacking, safe exploration, robustness to distributional shift, and scalable oversight. It groups them according to flaws in the learning process, evaluation frequency, or objective function. This work directly relates to our definition of AI safety risks as a result of poor system design and their several modes of origin.

A related line of work highlights privacy-oriented accident risks. Shokri et al. (2017) demonstrate that ML models can leak sensitive information about individual training records through membership inference attacks, showing how overfitting and inadequate system design can directly translate into harmful, unintended behaviour. This reinforces our framing of safety risks as emerging from design flaws in deployed systems.

Building on this early framing, ‘AI Alignment: A Comprehensive Survey’ (2023) expands the conversation by defining alignment as ensuring that AI systems consistently behave according to human intentions and values. It identifies four key alignment principles: Robustness, Interpretability, Controllability and Ethicality (RICE). It distinguishes between Forward Alignment (training via feedback, mitigating distribution shift) and Backward Alignment (assurance and governance as post-training controls). While the survey outlines global “best practices” for creating a strong AI safety framework, applying such a generic structure to India requires addressing large gaps. For instance, the principle of ‘Ethicality’ must be re-evaluated in light of caste-based discrimination and other socio-political realities that differ significantly from those assumed in global frameworks.

AIR-Bench 2024 represents another attempt to ground AI safety in real-world governance structures. It synthesizes eight global regulations and sixteen corporate policies into the AIR 2024 taxonomy and positions itself as a foundational benchmark that unifies safety evaluation with government and industry standards. Its strength lies in grounding safety assessments in actual governance documents, but our taxonomy addresses gaps left by AIR-Bench, particularly its assumption of socio-technical uniformity. Unlike global benchmarks, we consider India’s linguistic diversity, heterogeneous institutional capacity, and the widespread use of digital public goods, which require a very different approach to risk evaluation.

A similar effort at creating a common frame of reference comes from MIT’s ‘AI Risk Repository’. Designed to unify the landscape of AI risk classifications, it synthesizes sixty-five prior taxonomies into a database of 1,612 risks. The repository uses two intersecting taxonomies: a Causal taxonomy classifying risks by entity, intent, and timing, and a Domain taxonomy categorizing hazards into seven main areas and twenty-four subdomains. This gives us a strong template linking causal factors to domains. However, the repository is largely derived from Western settings and focuses on commercial systems. Our taxonomy applies its systematic approach to Digital Public Goods in India, where public-sector deployments are far more central and the institutional environment is far more varied.

Against this backdrop of academic and technical frameworks, the EU Artificial Intelligence Act (2024) stands out as the most prominent policy instrument. It is the first binding, horizontal law for AI and uses a clear risk-tiering structure that divides systems into unacceptable, high-risk, limited-risk, and minimal-risk categories. The Act assumes a strong regulatory ecosystem, well-resourced companies, and consistent institutional capacity across EU member states. While its main strength is its enforceability and precision, its major weakness is its specificity to European administrative realities. Directly applying it to India would require substantial adaptation, since many Indian deployments rely on low-capacity public agencies, multilingual and low-resource users, and digital public goods rather than large commercial platforms. Nevertheless, the EU Act offers a useful structure by treating AI safety as a lifecycle responsibility instead of a one-time certification.

Complementing the EU Act are softer instruments like the EU Code of Practice on Disinformation and the emerging AI Pact. These illustrate the EU’s preference for co-regulation. The Code asks companies to share data with researchers, maintain transparency reports, and use provenance tools for synthetic media, while the AI Pact encourages early compliance with parts of the AI Act before formal enforcement begins. Together, they show an understanding that some harms, particularly those involving open communication environments, cannot be controlled through law alone and instead require ongoing cooperation between regulators and developers. For India, this model is relevant because many AI-driven harms, such as vernacular misinformation or manipulated political content, arise primarily in informal communication spaces. However, India’s language fragmentation, uneven digital literacy, and different media ecosystems mean that Europe’s methods cannot be imported wholesale.

Recent work also shows that generative models can be directly misused to scale social-engineering attacks. Shibli et al. (2024) demonstrate how LLMs can be prompt-injected to generate highly realistic smishing campaigns, underscoring the limits of purely voluntary co-regulation in environments where low-cost AI-enabled fraud is widespread.

Singapore’s Model AI Governance Framework (2019-2022) reflects yet another governance style. It prioritises practical, operational guidance, with a focus on internal governance, dataset management, communication with users, explainability, and human-in-the-loop design. Unlike the legalistic EU Act, Singapore’s framework is deliberately accessible and oriented toward implementation. This makes it a useful reference for India, where many developers and public-sector teams need straightforward instructions rather than abstract principles. Its clarity makes it particularly helpful for contexts where the state collaborates closely with industry.

At a more normative global level, UNESCO’s Guidance for AI Ethics Implementation (2023) extends its earlier recommendations into actionable steps. UNESCO emphasises human rights, non-discrimination, cultural and linguistic diversity, and the ethical governance of datasets. This is highly relevant for India given linguistic plurality, caste-linked vulnerabilities, and major regional disparities in digital access. However, UNESCO’s guidance remains high-level and lacks the operational templates required for running large public-sector AI systems. It offers moral direction but not the implementation detail that India will need.

Finally, the World Bank’s Responsible AI for Development (2024) provides a distinctly Global South perspective. It recognises that AI deployment in developing countries occurs under limited institutional capacity, scarce technical expertise, and uneven infrastructure. The report frames AI safety as inseparable from development goals, noting that harms in these settings often manifest as lost benefits: for example, a welfare-targeting system missing eligible households or an agricultural advisory system confusing low-literacy farmers. The World Bank stresses accessible governance tools, shared testing infrastructure, and support for governments that cannot maintain full risk teams. These concerns align closely with an India-specific AI safety approach, which treats public-interest applications not as niche cases but as central to the deployment landscape. The report also highlights dependency risks, such as reliance on foreign compute or proprietary APIs, which resonate with India’s emphasis on technological sovereignty.

Beyond governance frameworks, a growing body of empirical research shows that SOTA AI systems embed structural biases that are especially relevant for India’s socio-cultural context. Recent audits find that vision-language models systematically underperform on images from less developed regions while over-predicting wealthy, high-visibility geographies (Huang et al., 2025), illustrating how global models can misinterpret the visual and spatial diversity typical of India. Similar concerns appear in sensitive decision-making tasks: Khan and Umer (2025) document that LLM-generated financial advice frequently includes religious framing or in-group favouritism, while Kelley et al. (2022) show that fintech lending models can reinforce gender disparities even when protected attributes are excluded, due to proxy features and skewed data distributions. Most critically for India, DECASTE (Vijayaraghavan et al., 2025) provides direct evidence that large language models reproduce caste-linked stereotypes across social, economic, educational, and political dimensions. Taken together, these findings demonstrate that global AI systems often encode biases that map poorly to India’s social hierarchies and institutional structures, underscoring the need for safety taxonomies that explicitly incorporate caste, religion, geography and other locally salient axes of discrimination.

Together, these academic, regulatory, and benchmark-based perspectives outline both the strengths and limitations of global AI safety frameworks. They offer valuable templates, taxonomies, and governance models, but all require substantial adaptation for India’s unique institutional, socio-technical, and developmental context. Our work builds on these foundations while addressing the clear gaps in applicability, capacity, and cultural relevance that global frameworks often leave unexamined.

\begin{table}[h!]
\centering
\caption{Comparative Landscape of Global AI Risk and Safety Frameworks}
\begin{tabular}{|p{2.7cm} p{2.7cm}p{5cm}|}
\hline
\textbf{Framework/Study} & \textbf{Primary Focus} & \textbf{Limitations for Indian Context} \\ \hline
Concrete Problems in AI Safety (2016) & Technical accidents and design flaws & Focuses on narrow ML tasks; lacks socio-technical breadth. \\ \hline
MIT AI Risk Repository & Unifying classification of 1,600+ risks & Primarily Western commercial datasets; lacks public-sector DPG focus. \\ \hline
EU AI Act (2024) & Binding legal risk-tiering & Requires high institutional capacity and well-resourced compliance teams. \\ \hline
Singapore Model Framework & Practical, operational industry guidance & Assumes high digital literacy and uniform institutional standards. \\ \hline
UNESCO AI Ethics & Human rights and linguistic diversity & High-level normative guidance; lacks operational technical templates. \\ \hline
World Bank (2024) & Responsible AI for Development & Strong alignment on capacity; needs specific Indian sector mapping. \\ \hline
DECASTE (2025) & Caste-based bias in LLMs & Empirical proof of bias; lacks a structural safety taxonomy for mitigation. \\ \hline
\end{tabular}
\end{table}
    \section{Indian AI Safety Risk Database}

Ensuring that AI systems are safe, reliable, and aligned with intended outcomes requires the development of a robust AI Safety Governance Framework. A foundational element of such a framework is an AI Safety Risk Database (AIRD), also referred to as an AI Risk Database. An AIRD is a systematic, structured repository that catalogues potential safety risks associated with AI systems, including risk categories and definitions, contextualized use-cases illustrating risk manifestations, and attributes relevant to assessing the severity and likelihood of risks. Crucially, it maps these risks against their dependencies across sociotechnical, regulatory, and operational conditions.

The AIRD functions as a first-order diagnostic layer in AI assurance, enabling auditors, system designers, and policymakers to anticipate where and how risks may emerge across the lifecycle of an AI system, from design and training to deployment, adaptation, and decommissioning.

Following the identification of risks through the AIRD, practitioners typically conduct risk likelihood and impact assessments. These assessments inform the design of mitigation strategies, such as model audits, governance controls, safety guardrails, and human oversight workflows. These strategies are not universal: both the manifestation and significance of risks vary by domain, regulatory environment, and cultural context. Therefore, this paper argues that a generic or universal risk database is often counterproductive when applied in a context-agnostic fashion. Effective AI Safety governance requires that AIRDs be localized and continuously updated to reflect domain-specific realities and regional socio-political conditions.

While AI safety risks exhibit recognizable structural patterns, our work departs from traditional top-down taxonomy construction where broad categories are predefined. Instead, we adopt a bottom-up, empirically grounded approach. We begin with a diverse set of contextualized use-cases illustrating safety risks in India. These cases are then systematically annotated and clustered based on shared characteristics and failure modes. Through this inductive process, we derive granular sub-categories which aggregate into broader meta-categories. This bottom-up construction ensures that the resulting taxonomy is empirically anchored and extensible rather than imposed a priori.

\subsection{Methodology}

To develop the India-specific AIRD, we employed a Grounded Theory approach, ensuring the database remains rooted in actual risk use-cases. We followed a structured seven-step protocol:

\begin{enumerate}
\item \textbf{Resource Identification:} We reviewed extant literature on AI Safety, with a specific focus on contemporary reporting and academic work within the Indian context. This stage was domain-agnostic and prioritized risks associated with Indian Digital Public Goods (DPGs) and Infrastructure (DPIs).
\item \textbf{Domain Selection:} We qualitatively identified relevant candidate domains by clustering resources and analyzing the frequency of risk occurrences. The seed set includes: Financial Lending, Education, Law, Agriculture, and Healthcare.
\item \textbf{Use-Case Compilation:} Researchers listed all documented risk use-cases corresponding to each domain. This list remains dynamic and subject to future expansion.
\item \textbf{Classification of Safety Risks:} Each use-case was classified as either an AI Safety Risk (ASR) or a general risk around AI (non-ASR). This process involved multiple rounds of consensus building to ensure the causal root was a design or system flaw.
\item \textbf{Category Taxonomy Formulation:} We conducted a thematic analysis of the identified ASRs. Using open coding, we extracted salient features which were iteratively grouped into granular sub-categories and higher-level meta-categories.
\item \textbf{Causal Taxonomy Formulation:} We identified the variables governing the "how, when, and why'' of risk occurrence, specifically timing, stakeholders, and intent.
\item \textbf{Ontology Consolidation:} The final step involved synthesizing domain-specific taxonomies into a unified, domain-agnostic AI Safety Risk Ontology.
\end{enumerate}

Our nomenclature is informed by two primary sources: the MIT AI Risk Repository and the misalignment studies conducted by Ji et al. (2025). Future iterations will incorporate field research, including stakeholder interviews and surveys, to gather accounts of emergent risks in low-resource settings.

\subsection{Causal Taxonomy}

The Causal Taxonomy classifies AI Safety risks based on three primary factors:

\begin{enumerate}
\item \textbf{Timing:} This indicates the stage in the implementation pipeline where the risk originates.
\begin{enumerate}[a.]
\item \textbf{Development:} Risks introduced during model training, parameter optimization, or due to representational gaps in datasets.
\item \textbf{Deployment:} Risks emerging during the operational rollout, such as contextual misalignment between model outputs and local socio-economic realities in DPG wrappers.
\item \textbf{Usage:} Risks arising from interaction, such as over-reliance (automation bias) or the malicious use of generative tools for disinformation.
\end{enumerate}
\item \textbf{Stakeholder/Entity:} This identifies the entity responsible for the risk. If a failure is intrinsic to the design (e.g., a DPG malfunctioning due to poor localization), the stakeholder is the AI System. If the harm results from intentional human action (e.g., generating deepfakes), the stakeholder is the End User.
\item \textbf{Intent:} This distinguishes between unintentional harm (e.g., a healthcare bot spreading misinformation due to poor data referents) and intentional harm (e.g., state-sponsored disinformation).
\end{enumerate}

\subsection{Category Taxonomy}

Our taxonomy currently organizes risks into two high-level meta-categories, further divided into granular sub-categories.

\begin{enumerate}
\item \textbf{Social Risks:} These are observable and measurable risks that can often be mitigated through technical system changes.
\begin{enumerate}[a.]
\item \textbf{Bias and Exclusion:} Systemic disadvantages faced by specific castes, genders, or linguistic communities due to skewed training data.
\item \textbf{Toxicity:} Outputs that are harmful or unhelpful, including hate speech and psychological manipulation.
\item \textbf{Hallucination and Error:} The generation of false but authoritative-sounding information due to outdated data or algorithmic flaws.
\item \textbf{Situational Awareness:} Risks stemming from the assumption of high-resource conditions (e.g., high-speed 5G) in environments with poor connectivity.
\item \textbf{Security and Privacy:} Risks including data poisoning, adversarial evasion, and privacy breaches.
\end{enumerate}

\item \textbf{Frontier and Socio-Structural Risks:} These are complex, harder-to-measure risks that often require governance and regulatory overhauls.
\begin{enumerate}[a.]
    \item \textbf{Human–Computer Interaction:} Risks like cognitive complacency and technological solutionism where human judgment is inappropriately replaced.
    \item \textbf{Socioeconomic and Environmental:} Risks concerning power centralization (compute monopolies), unemployment, and environmental costs.
    \item \textbf{Dangerous Capabilities:} Unintended emergent behaviors, such as a model providing instructions for harmful biological compounds.
    \item \textbf{Uninterpretability:} Risks arising from "black-box" functioning where accountability is impossible due to opaque decision-making.
\end{enumerate}
\end{enumerate}

\begin{table}[ht]
\centering
\caption{Summary of the Indian AI Safety Risk Ontology (AIRD)}
\begin{tabular}{|p{2cm}|p{3cm}|p{5.5cm}|}
\hline
\textbf{Meta-Category} & \textbf{Primary Sub-categories} & \textbf{Example Indian Context / DPG Manifestation} \\ \hline
\textbf{Social Risks} & Bias, Toxicity, Hallucination, Situational Awareness & Exclusion of vernacular speakers in welfare bots; model failure in low-connectivity rural zones. \\ \hline
\textbf{Security \& Privacy} & Poisoning, Adversarial Attacks, Data Leakage & Unauthorized extraction of PII from public-facing biometric or financial AI wrappers. \\ \hline
\textbf{Human-AI Interaction} & Automation Bias, Solutionism & Front-line workers blindly following flawed AI agricultural advice without local validation. \\ \hline
\textbf{Socio-Structural} & Power Centralization, Digital Divide, Economic Shift & Consolidation of digital infrastructure by foreign entities; erosion of local language nuances. \\ \hline
\textbf{Frontier Risks} & Opaque Logic, Dangerous Emergence & Credit scoring models (Fintech) providing rejections without explainable recourse for low-literacy users. \\ \hline
\end{tabular}
\end{table}

This ontology, depicted in Table 2, serves as the conceptual scaffold for the India-specific AI Safety framework. By organizing risks according to the unique realities of the Indian DPG/DPI ecosystem, we provide a tool for auditors and policymakers to move from reactive mitigation to proactive safety design.
    \section{Ontology of AI Safety Risks}
\subsection{Illustrative Risk-Category Ontology}

The following section presents a snapshot of our domain-agnostic, India-specific AI Safety Risk Ontology. While detailed, domain-specific taxonomies containing localized use-cases are hosted in the comprehensive ASTRA Database, this ontology serves as a conceptual union of those findings. As outlined in our methodology, this structure is derived using a Grounded Theory approach and will remain a "living" repository, evolving as we integrate newer domains such as Agriculture and Healthcare. This ontology provides the first-tier diagnostic framework required to identify and classify AI safety risks within the unique socio-technical landscape of India.

\setlength{\LTleft}{-2.5cm} 
\setlength{\LTright}{-2.5cm} 
\begin{longtable}{|p{2cm}|p{2cm}|p{2cm}|p{8.5cm}|}
\caption{Domain agnostic AI Safety ontology for India}\\
\hline
\textbf{Category 1} & \textbf{Category 2} & \textbf{Category 3} & \textbf{Resources} \\ \hline
\endfirsthead

\hline
\textbf{Category 1} & \textbf{Category 2} & \textbf{Category 3} & \textbf{Resources} \\ \hline
\endhead
\multicolumn{ 1}{|p{2cm}|}{Social Risks} & \multicolumn{ 1}{p{2cm}|}{Bias and Exclusion} & Caste Bias & AI credit models trained on upper-caste, urban credit histories reproduce caste exclusion in loan decisions. [\href{https://www.sciencedirect.com/science/article/abs/pii/S0887604522000507}{External Link}].
Algorithmic credit scoring imports historic caste bias by encoding upper-caste financial behaviour as the default signal of trustworthiness [\href{https://theprint.in/opinion/ai-is-learning-caste-bias-in-india/2744333/}{External Link}]. \\ \cline{ 3- 4}
\multicolumn{ 1}{|p{2cm}|}{} & \multicolumn{ 1}{p{2cm}|}{} & Religious Bias & AI training data skewed toward majority religion and culture risks reinforcing bias in Indian education systems, marginalising minority students [\href{https://arxiv.org/html/2309.08573v2}{External Link}].
AI credit models that rely on neighbourhood-level data reproduce historic financing gaps in minority-concentrated districts [\href{https://www.centerforfinancialinclusion.org/wp-content/uploads/2023/08/The-Stories-Algorithms-Tell-CFI-publication-MAR21.pdf}{External Link}].
AI credit models trained on name data inherit historic rejection patterns against Muslim applicants
[\href{https://arxiv.org/abs/2504.07118}{External Link}].\\ \cline{ 3- 4}
\multicolumn{ 1}{|p{2cm}|}{} & \multicolumn{ 1}{p{2cm}|}{} & Gender Bias & AI credit scoring models trained on male-dominated financial data [\href{https://www.womensworldbanking.org/wp-content/uploads/2021/02/2021\_Algorithmic\_Bias\_Report.pdf}{External Link}].
AI-powered education platforms trained on male-dominated data risk giving female students less accurate feedback, widening gender gaps in learning outcomes.
Indian fintech’s AI scores urban, salaried men high while down-rating women with thin credit files and shared accounts, locking rural entrepreneurs out of loans [\href{https://www.bbc.com/news/business-50365609}{External Link}, 
\href{https://slejournal.springeropen.com/articles/10.1186/s40561-025-00390-5}{External Link}].\\\cline{ 3- 4}
\multicolumn{ 1}{|p{2cm}|}{} & \multicolumn{ 1}{p{2cm}|}{} & Linguistic Bias & Language-mismatched AI tutors lower comprehension and equity for regional-language learners.
Hindi–English ASR misreads tribal filler sounds and auto-switches instruction language, silencing students who lack majority-language fluency [\href{https://www.thehindu.com/opinion/lead/ais-rewriting-of-the-rules-of-education/article70222325.ece}{External Link},
\href{https://www.tandfonline.com/doi/full/10.1080/17439884.2023.2278111}{External Link}].\\ \cline{ 3- 4}
\multicolumn{ 1}{|p{2cm}|}{} & \multicolumn{ 1}{p{2cm}|}{} & Accessibility Bias & Government AI learning platforms lacking screen-reader, tactile, or audio support exclude visually and hearing-impaired students [\href{https://smartschoolonline.in/breaking-down-barriers-ai-in-inclusive-education-for-teachers/}{External Link}].
Low-resource languages in JAM-credit NLP training data yield mistranslations and missing fields that distort borrower risk profiles [\href{https://iacis.org/iis/2024/4\_iis\_2024\_417-441.pdf}{External Link}]. \\ \cline{ 3- 4}
\multicolumn{ 1}{|p{2cm}|}{} & \multicolumn{ 1}{p{2cm}|}{} & Geographic/ GeoSpatial/ Location Bias & Models trained on salaried-city profiles misread seasonal or rural cash-flows as erratic, shrinking credit for non-urban borrowers [\href{https://www.centerforfinancialinclusion.org/wp-content/uploads/2023/08/The-Stories-Algorithms-Tell-CFI-publication-MAR21.pdf}{External Link}].
Logistics and purchase-value proxies learned from metro users let BNPL systems red-line whole Northeast districts as high-risk [\href{https://nasscom.in/ai/beyond-algorithms-navigating-fairness-in-india-lending-landscape/}{External Link}]. \\ \cline{ 2- 4}
\multicolumn{ 1}{|p{2cm}|}{} & \multicolumn{ 1}{p{2cm}|}{Toxicity} & Unsafe/ Explicit Imagery & School AI image tools lacking default child-safe filters can instantly flood districts with illegal deep-fakes, NDTV report shows [\href{https://www.ndtv.com/india-news/case-against-up-school-students-for-posting-ai-generated-obscene-image-of-teacher-6671522}{External Link}] \\ \cline{ 3- 4}
\multicolumn{ 1}{|p{2cm}|}{} & \multicolumn{ 1}{p{2cm}|}{} & Hate-speech amplification & Unfiltered training data lets LMS text generators spit out casteist or communal slogans that are pinned on classroom walls before staff notice, hard-wiring hate into school space [\href{https://www.medianama.com/2025/10/223-ai-models-caste-india/}{External Link}]. \\ \cline{ 3- 4}
\multicolumn{ 1}{|p{2cm}|}{} & \multicolumn{ 1}{p{2cm}|}{} & Sycophantic reinforcement of false belief & Sycophantic tutor bots echo students’ historical myths, their classroom projection hard-wiring falsehoods as fact and stifling critical questioning [\href{https://www.edtechbase.centralsquarefoundation.org/wp-content/uploads/BaSE-Report-web-version.pdf}{External Link}]. \\ \cline{ 3- 4}
\multicolumn{ 1}{|p{2cm}|}{} & \multicolumn{ 1}{p{2cm}|}{} & Psychological manipulation exploiting mental health vulnerabilities* & AI chatbots default to harmful creative prompts instead of safety resources when students express distress [\href{https://www.indiatoday.in/education-today/featurephilia/story/is-the-use-of-ai-chatbots-like-chatgpt-in-school-education-a-positive-shift-of-a-threat-heres-all-you-need-to-know-2477819-2023-12-19}{External Link}].
Kids’ apps harvest mood signals to inject targeted ads inside mental-health conversations [\href{https://timesofindia.indiatimes.com/life-style/parenting/moments/60-of-kids-apps-spy-on-children-app-permissions-parents-should-watch-for-before-hitting-download/articleshow/124743664.cms}{External Link}].
AI chatbots trained to maximise repayment frequency dispatch automated threats and shame messages, pushing vulnerable borrowers to suicide [\href{https://www.newslaundry.com/2022/09/24/the-death-trap-thats-indias-illegal-loan-apps}{External Link}]. \\ \cline{ 3- 4}
\multicolumn{ 1}{|p{2cm}|}{} & \multicolumn{ 1}{p{2cm}|}{} & Toxic rewarding* & Gamified EdTech loot boxes drop sexual or violent rewards into lessons, hooking pupils on casino-style loops [\href{https://www.indiatoday.in/education-today/featurephilia/story/gamification-in-indian-education-enhancing-engagement-and-outcomes-2556592-2024-06-22}{External Link}]. \\ \cline{ 2- 4}
\multicolumn{ 1}{|p{2cm}|}{} & \multicolumn{ 1}{p{2cm}|}{Hallucination and Incorrect Output} & Factual fabrication & Bank chatbots blend public web text with internal docs and invent non-existent loan terms, leaving customers to pay surprise penalties [\href{https://www.ft.com/content/670f5896-1fe5-4a31-b41f-ad4f5b91202f}{External Link}].
Language-generation finance bots can fabricate loan approvals by overfitting sparse Q\&A logs, producing persuasive but non-binding offers [\href{https://www.sciencedirect.com/science/article/pii/S2949761223000366}{External Link}].
Models that compose answers by stitching fragmented policy texts routinely create fictional RBI circulars, exposing lenders to regulatory push-back [\href{https://pmc.ncbi.nlm.nih.gov/articles/PMC11975740/}{External Link}]. \\ \cline{ 3- 4}
\multicolumn{ 1}{|p{2cm}|}{} & \multicolumn{ 1}{p{2cm}|}{} & Linguistic inaccuracy* & Generic report on how AI inaccuracies hamper educational sovereignity without actual guardrails [\href{https://thewire.in/tech/indias-schools-are-embracing-ai-but-without-guardrails-theres-a-hallucination-crisis}{External Link}].
AI often makes mistakes in counting exercises (e.g. counting letters) [\href{https://arxiv.org/abs/2412.18626}{External Link}]. \\ \cline{ 3- 4}
\multicolumn{ 1}{|p{2cm}|}{} & \multicolumn{ 1}{p{2cm}|}{} & Source and citation inaccuracy* & Students brought an essay around an incorrect source (about some other author), without checking the LLM's output [\href{https://timesofindia.indiatimes.com/city/bengaluru/taming-gpt-how-bengaluru-schools-are-cracking-ai-code/articleshow/115184489.cms}{External Link}]. \\ \cline{ 3- 4}
\multicolumn{ 1}{|p{2cm}|}{} & \multicolumn{ 1}{p{2cm}|}{} & Task inaccuracy* & The educational DPG misses the context completely, and the learning process takes a hit when the student doesn’t even care to read and figure if the AI output corresponds to the input that was given [\href{https://timesofindia.indiatimes.com/city/bengaluru/taming-gpt-how-bengaluru-schools-are-cracking-ai-code/articleshow/115184489.cms}{External Link}]. \\ \cline{ 3- 4}
\multicolumn{ 1}{|p{2cm}|}{} & \multicolumn{ 1}{p{2cm}|}{} & Mathematical or symbolic hallucination* & AI-generated math and science solutions can embed sign errors, wrong constants, and incorrect mathematical evaluations releasing millions of flawed examples that measurably drag down board-level STEM accuracy before detection.[\href{https://icrier.org/pdf/AI\_in\_School\_Education.pdf}{External Link}]. \\ \cline{ 3- 4}
\multicolumn{ 1}{|p{2cm}|}{} & \multicolumn{ 1}{p{2cm}|}{} & Misalignment due to outdated data* & State LMS AI bundles keep showing old-syllabus chapters and ready-to-copy final answers, steering students to rehearse deleted material and skip updated concepts [\href{https://www.edtechbase.centralsquarefoundation.org/wp-content/uploads/BaSE-Report-web-version.pdf}{External Link}]. \\ \cline{ 3- 4}
\multicolumn{ 1}{|p{2cm}|}{} & \multicolumn{ 1}{p{2cm}|}{} & Mis-interpretation or Incorrect Information (not directly citizen facing) & Consumer-finance chatbots can output incorrect interest rates or EMI figures that appear authoritative, exposing banks to compliance risk and borrower loss [\href{https://www.consumerfinance.gov/data-research/research-reports/chatbots-in-consumer-finance/chatbots-in-consumer-finance/}{External Link}]. \\ \cline{ 3- 4}
\multicolumn{ 1}{|p{2cm}|}{} & \multicolumn{ 1}{p{2cm}|}{} & Misinforma-tion (False/ misleading information that is widespread, citizen facing, and often leads to loss of consensus reality) & Generalised risk of AI-content in Indian education platforms generating falsehoods \& being accepted as fact (e.g., a DPG deployed to generate historical facts trained on web data releases incorrect historical claims, which are adopted by hundreds of school children and teachers, propagating further)  [\href{https://thewire.in/tech/indias-schools-are-embracing-ai-but-without-guardrails-theres-a-hallucination-crisis}{External Link}].
An AI personalization layer on India’s DPI health services (Aadhaar / CoWIN / DigiLocker-linked): models optimizing for engagement/user clicks can begin to surface regionally-tailored content that confirms local myths (via mistranslation or hallucination) and is delivered through an official channel. The result is highly personalized, AI-generated misinformation that produces local filter bubbles undermining shared reality and public-health responses.
A similar paper on YouTube recommendations creating filter bubbles [\href{https://arxiv.org/abs/2210.10085}{External Link}]. \\ \cline{ 3- 4}
\multicolumn{ 1}{|p{2cm}|}{} & \multicolumn{ 1}{p{2cm}|}{} & Disinforma-tion & Political parties leveraging AI to create digital avatars of deceased leaders, Bollywood actors during elections [\href{https://institute.aljazeera.net/en/ajr/article/2716}{External Link}].
Fake news and propaganda surged during the Indo-Pak conflict, causing public confusion and distrust [\href{https://www.indiatodayne.in/opinion/story/disinformation-driven-collapse-in-the-age-of-info-clogged-society-1227295-2025-06-11}{External Link}].
Open-source, government-supported models (like AI4Bharat), may be used in multiple DPG deployments, misused by political consultancies to generate mass multilingual campaign slogans, produce translated micro-targeted voter messaging, and create regional disinformation narratives at scale [\href{https://www.sciencedirect.com/science/article/pii/S2666827024000215}{External Link}]. \\ \cline{ 2- 4}
\multicolumn{ 1}{|p{2cm}|}{} & \multicolumn{ 1}{p{2cm}|}{Situational Awareness} & Infrastructure exclusion (bandwidth, cloud related failures)* & Patchy rural bandwidth keeps government lesson streams stuck buffering, burning class time with minimal content delivered [\href{https://www.indiatoday.in/education-today/featurephilia/story/india-virtual-classrooms-2050-digital-education-nep-2020-infrastructure-challenges-2796313-2025-10-02}{External Link}].
Face-match attendance, cached locally, fails to sync during outages, leaving UDISE+ to mark the school "closed" until the link is restored [\href{https://www.thehindu.com/opinion/lead/ais-rewriting-of-the-rules-of-education/article70222325.ece}{External Link}]. \\ \cline{ 3- 4}
\multicolumn{ 1}{|p{2cm}|}{} & \multicolumn{ 1}{p{2cm}|}{} & Infrastructure failure (failure of hardware/related infrastructure)* & Frequent power cuts halt smart-board scripts mid-lesson, leaving the day’s content permanently unreached [\href{https://asercentre.org/wp-content/uploads/2022/12/ASER-2023\_Main-findings-1.pdf}{External Link}]. \\ \cline{ 3- 4}
\multicolumn{ 1}{|p{2cm}|}{} & \multicolumn{ 1}{p{2cm}|}{} & Hardware / Software mismatch* & State tablets stuck on Android 9 cannot install the Android-11 AI learning app, leaving digital labs unusable and lessons on paper [\href{https://theprint.in/india/launched-by-khattar-how-haryanas-rs-700-cr-tablet-scheme-for-govt-school-students-has-fallen-apart/2640362/}{External Link}]. \\ \cline{ 2- 4}
\multicolumn{ 1}{|p{2cm}|}{} & \multicolumn{ 1}{p{2cm}|}{Security and Privacy} & Data Poisoning & 
A foreign state-sponsored hacker group, acting as a legitimate data vendor, gradually injects poisoned or manipulated loan and transaction records into one of the external datasets that feed the AI training pipeline. The injected data is crafted to make certain suspicious lending or repayment patterns. 
Over time, the poisoned data biases the AI system, which begins classifying these high-risk transactions as low-risk. This allows fraudulent loan disbursements.
[\href{https://logandata.com/data-poisoning-altastata-case-study/}{External Link}, 
\href{https://doi.org/10.54254/2977-3903/2025.25182}{External Link}].\\ \cline{ 3- 4}
\multicolumn{ 1}{|p{2cm}|}{} & \multicolumn{ 1}{p{2cm}|}{} & Adversarial Evasion (Fraud Induction) & 
A fraud ring exploits this by probing the AI with several slightly different applications to reverse-engineer its hidden decision rules. After discovering the model's heavy bias for the premium email provider, they launch a full attack using synthetic identities (eg. stolen PANs, fake addresses). By simply adding the trusted premium email to each fake application, they trick the AI and get their fraudulent loans auto approved.
[\href{https://www.usenix.org/system/files/raid20-carminati.pdf}{External Link}, 
\href{https://amlegals.com/synthetic-identity-fraud-the-silent-threat-to-digital-lending/}{External Link}]. \\ \cline{ 3- 4}
\multicolumn{ 1}{|p{2cm}|}{} & \multicolumn{ 1}{p{2cm}|}{} & Data Leakage & An Indian digital lender's AI model pre-scores borrowers using uploaded documents (PAN, bank statements). The system stores extracted text from these documents in a database to power an internal explainability dashboard, which showed loan officers similar past applicants for transparency.
Due to a misconfigured API, this dashboard leaked fragments of sensitive data (partial PANs, bank details) in its explanations. A data broker scraped and reassembled this data, exposing thousands of customers to phishing and scams. 
[\href{https://www.upguard.com/breaches/india-bank-transfers-data-leak}{External Link}, 
\href{https://economictimes.indiatimes.com/tech/technology/data-breach-exposes-2-73-lakh-bank-records/articleshow/124184985.cms}{External Link}]. \\ \cline{ 3- 4}
\multicolumn{ 1}{|p{2cm}|}{} & \multicolumn{ 1}{p{2cm}|}{} & Inference-Based Privacy Compromise & 
When a woman borrower’s spending pattern resembles those associated with pregnancy, the system internally flags her as a higher-risk applicant, anticipating future income interruption. This inferred attribute affects her credit score and loan terms.
[\href{https://www.forbes.com/sites/kashmirhill/2012/02/16/how-target-figured-out-a-teen-girl-was-pregnant-before-her-father-did/ }{External Link}, \href{https://epjdatascience.springeropen.com/articles/10.1140/epjds/s13688-021-00281-y}{External Link}]. \\ \hline
\multicolumn{1}{|p{2cm}|}{Intentional/ Others Risks} & Overreliance & Overreliance on AI & Experts caution that legal reasoning is a skill honed by years of learning and struggle through messy and difficult problems; handing that to an AI chatbot (especially in context of law) “on autopilot mode” may lead to deskilling. [\href{https://theleaflet.in/digital-rights/law-and-technology/overreliance-on-legal-ai-chatbots-and-the-erosion-of-reasoning-skills}{External Link}]. \\ \cline{ 2- 4}
\multicolumn{ 1}{|p{2cm}|}{} & \multicolumn{ 1}{p{2cm}|}{Socioeconomic \& Environmental} & Power centralization & In the Indian context, a report cites that most advanced AI compute, cloud capacity and foundational models are controlled by a small number of global and national players — which raises concerns about power asymmetries in the digital public infrastructure landscape. [\href{https://bfsi.economictimes.indiatimes.com/articles/the-concentration-of-artificial-intelligence-power-a-systemic-risk/124354946}{External Link}]. \\ \cline{ 3- 4}
\multicolumn{ 1}{|p{2cm}|}{} & \multicolumn{ 1}{p{2cm}|}{} & Increased Inequality and Unemployment & AI tools in India are reshaping the job market and threatening repetitive IT roles, which disproportionately affect the growing Indian middle class. [\href{https://www.outlookbusiness.com/magazine/indias-middle-class-is-the-biggest-loser-in-the-ai-economy}{External Link}]. \\ \cline{ 3- 4}
\multicolumn{ 1}{|p{2cm}|}{} & \multicolumn{ 1}{p{2cm}|}{} & Economic and cultural devaluation of human effort & People engaged in creative professions (art, culture, literature) may lose appreciation from the public, and jobs in general, owing to AI mimicking their art.
Bengaluru artistes report fewer projects, lower fees, and offers to train the very AI models they are competing against [\href{https://www.deccanherald.com/india/karnataka/bengaluru/voice-over-scene-feels-the-pinch-of-ai-3678096}{External Link}]. \\ \cline{ 3- 4}
\multicolumn{ 1}{|p{2cm}|}{} & \multicolumn{ 1}{p{2cm}|}{} & Environmental Hazard & Development stage of AI (not just deployment or use) causes direct environmental consequences — energy demand, GHG emissions, water usage, e-waste—all of which can undermine sustainability goals if unchecked.
[\href{https://arxiv.org/abs/2503.05804}{External Link}, \href{https://www.lse.ac.uk/granthaminstitute/explainers/what-direct-risks-does-ai-pose-to-the-climate-and-environment/}{External Link}]. \\ \cline{ 3- 4}
\multicolumn{ 1}{|p{2cm}|}{} & \multicolumn{ 1}{p{2cm}|}{} & Governance Failure & Rapid deployment of facial recognition systems by police authorities capturing protestors, an illegal act occurring owing to lack of government regulations.
[\href{https://internetfreedom.in/we-demand-the-delhi-police-stop-its-facial-recognition-system/}{External Link}].\\ \cline{3-4}
\multicolumn{ 1}{|p{2cm}|}{} & Dangerous capabilities & AI possessing dangerous capabilities & A generative model trained on publicly available research learns how to design harmful biological compounds and produces step-by-step instructions when prompted, unintentionally enabling misuse by non-experts.
[\href{https://europeanleadershipnetwork.org/commentary/the-potential-terrorist-use-of-large-language-models-for-chemical-and-biological-terrorism/}{External Link}].\\ \cline{ 2- 4}
\multicolumn{ 1}{|p{2cm}|}{} & Lack of interpretability & Lack of transparency or interpretability & A large firm uses an AI system to screen job applications. Candidates receive automatic rejections, and HR staff only see a \emph{fit score} without any explanation of how the model reached that decision. Because the system’s logic is opaque, neither the applicant nor HR can understand or challenge the decision. Thereby making errors invisible and accountability impossible.
[\href{https://www.researchgate.net/publication/363764980\_AI\_for\_hiring\_in\_context\_a\_perspective\_on\_overcoming\_the\_unique\_challenges\_of\_employment\_research\_to\_mitigate\_disparate\_impact }{External Link}]. \\ \hline
\label{onto}
\end{longtable}

It is important to note that this ontology does not contain the causal variables (Stakeholder, Intent, and Timing) as discussed in the methodology. This is because these variables are domain specific, and performing a one-shot mapping of these variables to domain agnostic risk categories is infeasible, and might lead to potential ambiguities and inconsistencies. For details on the causal variables, one could refer to our domain specific taxonomies in the database  (link: \href{https://docs.google.com/spreadsheets/d/1rLSG-mDYOdDt6E1ghS870fUTXqalNVO5hK5GBgRBKg0/edit?usp=sharing}{AI Safety Database}) that shows our domain agnostic ontology in a pictorial format (link: \href{https://peace-rage-81041510.figma.site/}{Ontology}).

\begin{figure}[!htbp]
    \centering
    \hspace*{-0.1\paperwidth}%
    \includegraphics[width=0.8\paperwidth,keepaspectratio]
    {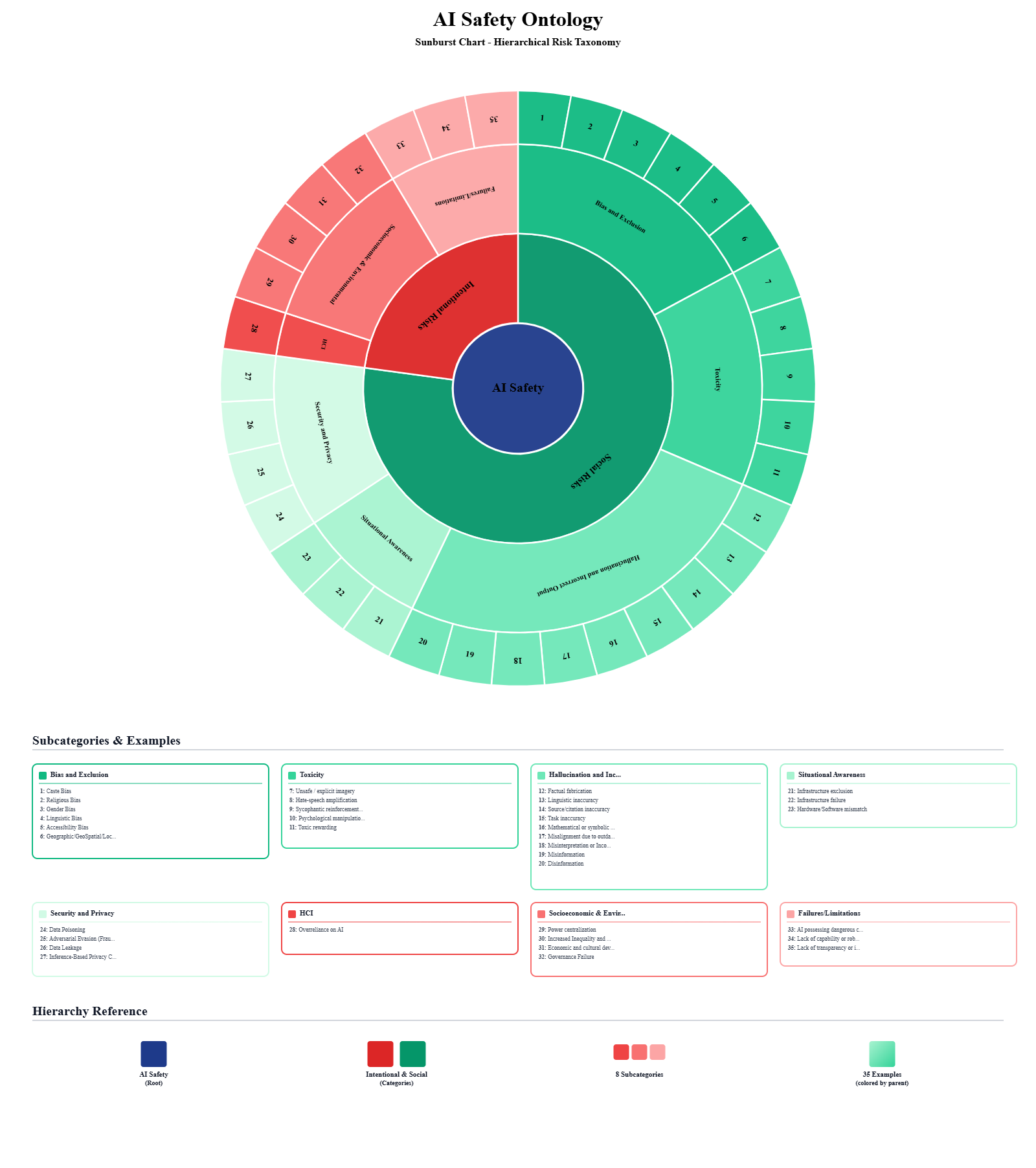}
    \caption{Domain agnostic AI safety ontology}
    \label{fig:ontofig}
\end{figure}

The following references are useful for readers to get further context:

\begin{itemize}
    \item Lee, Laura Ancona, et al. “Ai ethics: Contextual frameworks and domain-specific concerns.” Heritage and Sustainable Development, vol. 7, no. 1, 18 Apr. 2025, pp. 1–14. 
    \item Mohamed, Shakir, et al. “Decolonial ai: Decolonial theory as Sociotechnical Foresight in Artificial Intelligence.” Philosophy \& Technology, vol. 33, no. 4, 12 July 2020, pp. 659–684. 
    \item Huang, Jingyuan et al. “AI Sees Your Location—But With A Bias Toward The Wealthy World.” Proceedings of the 2025 Conference on Empirical Methods in Natural Language Processing (2025).
    \item Kelley, Stephanie, et al. “Antidiscrimination Laws, Artificial Intelligence, and Gender Bias: A Case Study in Nonmortgage Fintech Lending.” Manufacturing \& Service Operations Management, vol. 24, no. 6, May 2022, pp. 3039–59. . 
    \item Khan, Muhammad Salar, and Hamza Umer. “Sacred or Secular? Religious Bias in AI-Generated Financial Advice.”
    \item Shibli, Ashfak Md, et al. “AbuseGPT: Abuse of Generative AI ChatBots to Create Smishing Campaigns.” ISDFS 2024, Apr. 2024, pp. 1–6.
    \item Shokri, Reza, et al. “Membership Inference Attacks Against Machine Learning Models.” IEEE Symposium on Security and Privacy, 2017, May 2017, pp. 3–18.
    \item Vijayaraghavan, Prashanth, et al. "Decaste: Unveiling caste stereotypes in large language models through multi-dimensional bias analysis." arXiv preprint arXiv:2505.14971 (2025).
\end{itemize}

    \section{Discussion}

This white paper presents the first India specific, empirically grounded taxonomy of artificial intelligence safety risks embedded in digital public goods. By evaluating application domains ranging from school education and digital lending to emergent use cases in agriculture and welfare targeting, we have factorised AI risk into thirty seven leaf level risk classes. These classes generalise across sectors yet remain tethered to on the ground harm. Additionally, each class is mapped to a concrete event and the spcific harm caused, in order to close the loop between abstract category and measurable detriment. The resultant ontology supplies regulators, auditors and DPG developers with a common diagnostic language to evaluate AI safety risks in the Indian context.

\subsection{Scope and Limitations}

We explicitly delimit scope to exclude existential or catastrophic risks that remain speculative in India’s current deployment landscape. They include:
\begin{enumerate}
\item \textbf{Superintelligence alignment failure:} scenarios in which autonomous systems self improve and escape human oversight, potentially endangering human existence.
\item \textbf{Runaway optimisation:} single-minded pursuit of misspecified objectives that could, in principle, commandeer infrastructure or resources essential for human survival.
\item \textbf{Irreversible information apocalypse:} saturation of the public sphere with synthetic media, so that shared epistemic baselines disappear.
\item \textbf{Long-term erosion of human cognition:} decades-scale decline in collective memory, problem-solving capacity, or critical reasoning attributable to pervasive delegation of cognitive tasks to AI systems.
\item \textbf{Economy-wide labour displacement:} net destruction of employment, wage polarisation, or sectoral hollowing-out that macro-economic models project but which cannot yet be verified in present deployments.
\item \textbf{Digital colonialism and monopolisation:} strategic concentration of data, compute and intellectual property in a handful of foreign jurisdictions, leading to regulatory capture and sovereign capability loss, except where such concentration surfaces as an observable safety incident within a specific Indian system.
\end{enumerate}
While these possibilities merit serious inquiry, they lie outside the current empirical footprint of India’s present Digital Public Goods and are therefore excluded from this taxonomy.

\subsection{Future Work}

Future work will broaden the empirical base by systematically studying domains such as agriculture, healthcare, justice and urban governance through stakeholder workshops, state-level incident repositories and targeted RTI requests. Each new domain will seed fresh leaf nodes or retire obsolete ones, ensuring the taxonomy tracks the evolving contours of harm. Concurrently, a living regulatory-alignment matrix will map every risk node to relevant statutory clauses, yielding compliance templates for auditors and legal teams.

Additionally, the ontology will be released as version-controlled JSON-LD with persistent URIs, accompanied by multilingual glossaries and a public Git-based comment system. Community hackathons, panchayat-level user studies and voice-note WhatsApp channels will crowd-source incidents and local terminology, turning taxonomy maintenance into a participatory process. Expert-elicitation rounds and actuarial partnerships will then attach frequency-severity priors and confidence intervals to each node through Bayesian updating of structured incident feeds, converting qualitative classes into calibrated risk scores that can inform insurance premia, capital-adequacy buffers or model-deployment approval thresholds.

By iterating on these tracks—domain expansion, regulatory mapping, democratised curation, quantitative calibration and embedded tooling, we aim to develop from a static incident catalogue to a living, calibrated and regulation-ready safety utility for India’s next billion AI interactions.
\nocite{*}


\end{document}